\documentclass[12pt]{article}
\usepackage{amsbsy,graphicx}

\newcommand{\beq}{\begin{equation}}
\newcommand{\eeq}{\end{equation}}
\newcommand{\bea}{\begin{eqnarray*}}
\newcommand{\eea}{\end{eqnarray*}}
\newcommand{\beaq}{\begin{eqnarray}}
\newcommand{\eeaq}{\end{eqnarray}}
\begin{document}
\begin{flushright} KIAS-P03014\\cond-mat/0302282\end{flushright}
\vspace{10mm}
\centerline{\Large \bf Non-commutative field theory approach to}
\centerline{\Large \bf two-dimensional vortex liquid system}
\vskip 1cm
\centerline{\large Kyungsun Moon$^{1}$, Vincent Pasquier$^{2}$,  
Chaiho Rim $^{3}$, and Joonhyun Yeo$^{4}$} 
\vskip 1cm

\centerline{\it $^{1}$ Institute of Physics and Applied Physics, 
Yonsei University}
\centerline{\it Seoul 120-749, Korea}
\centerline{\it kmoon@phya.yonsei.ac.kr}
\vskip .5cm
\centerline{\it $^{2}$ Service de Physique Theorique
CEA/Saclay}
\centerline{\it  Orme des Merisiers 
F-91191 Gif-sur-Yvette Cedex, FRANCE}
\centerline{\it pasquier@spht.saclay.cea.fr}
\vskip .5cm
\centerline{\it $^{3}$ Department of Physics, Chonbuk National University}
\centerline{\it Chonju 561-756, Korea}
\centerline{\it rim@mail.chonbuk.ac.kr}
\vskip .5cm
\centerline{\it $^{4}$ Department of Physics, Konkuk University}
\centerline{\it Seoul 143-701, Korea}
\centerline{\it jhyeo@konkuk.ac.kr}
\vskip 2cm
\centerline{\bf Abstract}
\vskip 0.5cm
\noindent
We investigate the non-commutative (NC) field theory approach 
to the vortex liquid system 
restricted to the lowest Landau level (LLL) approximation.
NC field theory effectively 
takes care of the phase space reduction 
of the LLL physics in a $\star$-product form
and introduces a new gauge invariant form of
a quartic potential of the order parameter
in the Ginzburg-Landau (GL) free energy.
This new quartic interaction coupling term
has a non-trivial 
equivalence relation with that 
obtained by Br\'ezin, Nelson and Thiaville
in the usual GL framework. The consequence of the 
equivalence is discussed. 

\newpage

{\it 1. Introduction:} 
In the presence of a strong magnetic field, 
condensed matter systems of charged particles 
are often characterized by the lowest
Landau level (LLL) physics. 
In high-Tc superconductors,  for example, 
thermal fluctuations 
are much more effective than they are in conventional
low-Tc superconductors 
because of strong anisotropy, high temperature,
and short coherence length \cite{blatter}. 
A large portion of the field-temperature phase diagram
of a high-Tc superconductor is occupied
by the vortex liquid phase resulting from the melting
of the Abrikosov vortex lattice. 
Since, 
only fluctuations in the LLL order parameter are important
near $H_{c2}(T)$, the LLL approximation,
where the higher Landau modes are neglected all together,
has been widely used to study the vortex liquid phase
in high-Tc superconductors as well.
The higher Landau modes 
effectively renormalize the phenomenological parameters 
in the LLL theory, and the quantitative studies \cite{tes} 
of this effect show that the LLL approximation 
is valid over a wide range of phase diagram below $H_{c2}(T)$.

The vortex system 
was orignally studied by Abrikosov 
\cite{abrikosov} 
using LLL Ginzburg-Landau (GL) theory.
Br\'ezin, Nelson and Thiaville
(BNT) \cite{bnt} first
studied the fluctuation effects near the upper critical field
$H_{c2}(T)$ in type-II superconductors
using the functional renormalization group (RG)
on the LLL GL theory
and found that the fluctuations
drive the phase transition into first order.

On the other hand, the projection to the LLL  
completely quenches  
the kinetic energy 
for the two-dimensional system
and induces non-commutativity between two otherwise
independent coordinate variables 
just as it appears in matrix
multiplications \cite{nc}.  
Due to this non-commuting nature between coordinates
special care has to be taken to analyze LLL.

A useful tool 
to deal with these systems is the 
non-commutative (NC) field theory \cite{ac,ws},
which easily incorporates 
the phase space reduction of the system. 
The main advantage of NC field theory
is that one can use the ordinary field
theoretical technique used for commuting coordinates, 
but with field multiplication replaced 
by the $\star$-product. 
In fact, NC field appears in many different contexts
from quantum gravity at plank scale
and string theory \cite{gr}
to condensed matter system.
Recent studies of NC real scalar field theories 
show many interesting properties
such as non-commutative solitons 
\cite{soliton}
and phase structure \cite{sondhi}.

The quantum Hall (QH) system is considered as 
the most exemplary case of NC field theory 
applicable in condensed matter physics.
NC field approach is done for the system with 
strong magnetic field in \cite{b-s}.
The rigid fluid motion is described 
in terms of NC Chern-Simons field theory 
to understand quantization of the filling factor
of the QH system \cite{nccs} 
and the relation between NC $U(1)$ gauge theory
and fluid mechanics is investigated in \cite{ncfluid}.
In addition, the skyrmion excitations 
in QH system \cite{skyrmion}
have been studied based on the NC nature
of fermionic variables \cite{pasq,Lmr}.
In this paper we will consider 
the vortex liquid system 
near the upper critical field $H_{c2}$
in a high-Tc superconductor 
in a magnetic field.
This vortex system is another example of LLL physics, 
which can be explored using the NC field theory.

The two-dimensional superconductor in a uniform magnetic field  
is effectively described in terms of GL free energy, 
\beq
F[\Psi,\Psi^\dagger]
 = \int d^2 x  \left( \frac 1{2m} 
\Big| \Big( -i\hbar  \vec \nabla 
- \frac {e^*}c \vec A \Big) \Psi \Big|^2 
 + \alpha |\Psi (\vec x) |^2 
+ \frac\beta 2 |\Psi (\vec x) |^4 \right)\,,
\label{gl}
\eeq
where $\Psi$ is an order parameter  representing 
Cooper pair condensate wave-function and  $\alpha$, $\beta$ 
and  $m$ are phenomenological parameters, and $e^*=2e$.  
The vector potential is given as $\vec A = B ( -y/2, x/2)$ 
in the symmetric gauge.  
We consider the case where the order parameter 
and its fluctuations are restricted to the LLL. 
This is believed to be a good approximation 
over a wide range of the phase diagram 
below the upper critical field.  

The LLL GL free energy \cite{bnt} is put as $ H = H_2 + H_4 $
with
\beq 
H_2 = \alpha_2 \int d^2 x \,\, 
|\Psi_{\rm LLL} (\vec x) |^2\,\,, \qquad\qquad  
H_4 = \frac{\alpha_4}{2}  
\int d^2 x\,\, |\Psi_{\rm LLL} (\vec x) |^4 , \label{H}
\eeq 
where 
$\alpha_2 =\alpha+\hbar\omega_c/2$, 
$\alpha_4=\beta$, and $\omega_c = e^*B/(mc)$ 
is the cyclotron frequency.  The magnetic length 
$\sqrt{\hbar c /e^* B}$, 
$\hbar$ and $c$ will be set to 1 in the following.
The LLL order parameters are given as  
$ \Psi_{\rm LLL} (\zeta, \bar \zeta)  
= \phi(\zeta) \,e^{-\zeta \bar \zeta /2} $ 
where $\phi(\zeta)$ is an arbitrary holomorphic function
and $\zeta =(x+iy)/\sqrt2 $ ($\bar \zeta =(x-iy)/\sqrt2 $)
is the holomorphic coordinates (anti-holomorphic ones).
We will omit the subscript LLL from now on.  

For a systematic evaluation of the partition function,  
the order parameter is conveniently written in the
momentum space as 
\beq 
\Psi(\zeta,\bar\zeta)  
=\phi(\zeta) e^{-\frac 1 2 \zeta\bar\zeta}  
=\int\frac{d^2 k}{(2\pi)^2}\, 
\widetilde\Psi (k,\bar k)\, \exp
\left(\frac {i} 2 (k\bar\zeta+\bar k \zeta)\right)\,. 
\label{lll} 
\eeq 
where 
$\widetilde\Psi(k,\bar k) 
= 2\pi \exp(-k\bar k/2)\,
\phi(\frac 2 i \frac\partial {\partial\bar k})$.  
After integration by parts  one can obtain 
an equivalent form 
$\widetilde\Psi(k,\bar k)
=2\pi \exp(-\frac 1 2 k\bar k)\phi(-ik) $
and $\phi(-ik)$ is the coordinate holomorphic function
$\phi(\zeta)$ with $\zeta$ replaced by $-ik$.    

As first noted by BNT \cite{bnt}, the
renormalization in the LLL can be summarized in 
an effective gauge invariant quartic term as  
\bea 
H^{\rm BNT}_4 
=  \alpha_4\int  d^2 \zeta_1 d^2 \zeta_2 \,\,  
|\Psi(\zeta_1 ,\bar\zeta_1 )|^2 \,\, 
g^{\rm BNT}( \zeta_1-\zeta_2,\bar \zeta_1-\bar \zeta_2) \,\, 
|\Psi(\zeta_2 ,\bar\zeta_2 )|^2 
\eea
The Fourier transform representation of the quartic term
is given as
\beaq
H^{\rm BNT}_4 
&=& \alpha_4 \int  \left[ \prod_{i=1}^4 
\frac{d^2 p_i}{(2\pi)^2} \right] \, 
(2\pi)^2 \delta^{(2)}(\vec p_1 +\vec p_2 -\vec p_3 -\vec p_4 ) \,\, 
\tilde g^{\rm BNT}(p_3-p_1,\bar p_3-\bar p_1)
\nonumber \\ 
&&\quad \times(2\pi)^4 e^{-\frac{1}{2}\sum_i p_i \bar p_i }\,\, 
\phi(-ip_1)\,\phi(-ip_2)\,\phi^\dagger(ip_3)\,\phi^\dagger(ip_4)\,.  
\label{hbnt} 
\eeaq 
Here  $\tilde g^{\rm BNT}(k, \bar k) $
is the Fourier transform of 
$ g^{\rm BNT}(\zeta, \bar \zeta) $,
\bea 
g^{\rm BNT}( \zeta,\bar{\zeta})
=\int\frac{d^2 \vec k}{(2\pi)^2} \,\, 
\widetilde g^{\rm BNT} (k,\bar{k})\, 
\exp(\frac{i}{2}(k\bar\zeta +\bar k \zeta))\,.
\eea 
$\tilde g^{\rm BNT}(k, \bar k) $
takes into account  the renormalization of the quartic term 
starting from the bare value  
$\tilde g_0^{\rm BNT} (k, \bar k) = 1$.
As was demonstrated in \cite{mn}
the Fourier transformed representation of the 
kernel $g^{\rm BNT}(\zeta,\bar\zeta)$
has many advantages in perturbative calculation.
The Fourier transform 
is also directly related to physical quantities
describing the vortex liquid
such as the structure factor \cite{ym}.

\vskip 0.5cm
{\it 2. NC effective theory:}
The effective the LLL theory 
can be reformulated using NC complex bosonic field theory,
using the coherent state representation, 
\begin{equation}
\langle \zeta | l \rangle
 = \frac1{\sqrt{2 \pi l!}} \zeta^l \exp(-\frac12 \bar\zeta \zeta) ,
\end{equation}
where $|l \rangle$ is the angular momentum state.
The coherent states consist of the 
(over-) complete set of the LLL system,
\begin{equation}
<\zeta |\zeta ' > = \sum_l <\zeta |l> <l|\zeta '> 
=\frac 1{2\pi}  e^{- \frac {\zeta \bar \zeta}{2} 
- \frac{\zeta'  \bar \zeta ' }{2} 
+\zeta \bar \zeta ' }. 
\end{equation}

The above innocent looking description of the LLL 
in terms of the coherent states shows an essential feature of the LLL. 
Note that the coherent state description 
of the one-dimensional harmonic oscillator system 
comes from the minimal uncertainty wave packet, 
between the coordinate $x$ and the momentum $p_x$. 
In the LLL, it is not the coordinate and the momentum  
that do not commute, but the two coordinates $x$ and $y$.  
The non-commuting property of the  two-dimensional coordinates  
$x $ and $y$, or $\zeta$ and $\bar \zeta$ is main result 
of the phase space reduction due to the strong magnetic field.   

Therefore, the two coordinates $\zeta$ and $\bar \zeta$ 
are to be treated as non-commuting operators.
 This raises the ordering problem of coordinates and results 
in the awkward calculus of the  analysis \cite{nc}.   
To circumvent this inconvenient manipulation, 
one may introduce the non-commuting operators  
and the $\star$-product of the corresponding functions \cite{gr}. 
We regard $x$ and $y$ 
as {\it ordinary commuting} coordinates 
but instead encode the ordering information 
into the wave-function utilizing the $\star$-product. 

Suppose a function $f(\zeta, \bar \zeta)$ is given  
as a coherent state expectation value of an operator $O_f$: 
$f(\zeta, \bar \zeta) = <\zeta| O_f |\zeta>$,
then $\star$-product of the two functions are defined   
as the Moyal product 
\beq 
f\star g (\zeta, \bar\zeta) 
= \exp(\partial_{\bar\zeta} \partial_{\zeta'} 
- \partial_{\zeta} \partial_{\bar\zeta'})  
f(\zeta, \bar\zeta) g(\zeta', \bar\zeta')
\left.\right|_{\zeta = \zeta'}
\eeq 
in consistent with operator product representation, 
\beq 
<\zeta| \hat O_f  \hat O_g |\zeta> 
\equiv <\zeta|  \hat O_{f\star g}|\zeta>\,. 
\eeq 
The hatted operator $\hat O_f $ can be different 
from the unhatted one  $O_f $ 
by the amount of the normal ordering. 

The quadratic term GL free energy
is reformulated  as 
\beq 
K_2 = 2\alpha_2 \int d^2 \zeta\, 
\langle \zeta| \widehat\Psi^\dagger_{op} \, 
\widehat\Psi_{op} |\zeta \rangle  
=2\alpha_2\int d^2\zeta \,\,  
\Psi^\dagger\star\Psi (\zeta,\bar\zeta)  
= 2\alpha_2\int d^2\zeta \,\, |\Psi(\zeta,\bar\zeta) |^2\,.
 \label{k2} 
\eeq 
This quadractic part of the two Hamiltonians $H_2$ and $K_2$ 
are made equal thanks to the hatted operator 
and hence by the  nature of the Moyal product: 
Integration of Moyal product of two functions 
is the same as that of the ordinary product.  

The NC quartic term is written as 
\beaq
K_4^{\rm NC}&=&
\alpha_4\int d^2\vec\zeta \, d^2 \vec\zeta'\,\,
\Big[\Psi\star\Psi^\dag(\zeta,\bar\zeta)\Big]\,
g^{\rm NC} (\zeta-\zeta',\bar\zeta-\bar\zeta')\,
\Big[\Psi\star\Psi^\dag(\zeta',\bar\zeta')\Big]
\nonumber\\
&=& \alpha_4 \int  \left[ \prod_{i=1}^4
\frac{d^2 p_i}{(2\pi)^2} \right] \, 
(2\pi)^2 \delta^{(2)}(\vec p_1 +\vec p_2 -\vec p_3 -\vec p_4 ) \,\, 
\tilde g^{\rm NC}(p_3-p_1,\bar p_3-\bar p_1)
\nonumber \\ 
&&\quad \times v(\{p_i,\bar p_i\}) \,
(2\pi)^4 e^{-\frac{1}{2}\sum_i p_i \bar p_i }\,\,
\phi(-ip_1)\,\phi(-ip_2)\,\phi^\dagger(ip_3)\,\phi^\dagger(ip_4)\,, 
 \label{k4} 
\eeaq
where the Fourier transformed $\tilde g^{\rm NC} (k,\bar k)$ 
is the renormalized one 
with the  bare value $\tilde g^{\rm NC} _0(k,\bar k)=1$. 
This quartic term includes 
a new phase factor 
$v(\{p_i,\bar p_i\})$,
\beaq
v (\{ p_i ,\bar p_i\})
 &=& e^{\frac{1}{8}
\left[ (p_1 \bar p_3 - \bar p_1 p_3)
+(p_2 \bar p_4 -\bar p_2 \bar p_4 ) \right]}
 +(p_3 \leftrightarrow p_4)
\nonumber\\
&=& e^{ \frac{i}{4}
\left[ - (\vec p_1 \times \vec p_3)
-(\vec p_2 \times \vec p_4 ) \right]}
+(\vec p_3 \leftrightarrow \vec p_4)\,.
\label{v}
\eeaq

The quartic term (\ref{k4}) 
is manifestly different from BNT construction 
(\ref{hbnt}), 
even though both of them are gauge invariant.
The gauge transformation is represented as 
a translation in the vector potential,
$\vec{A} (\vec r) \to\vec{A}(\vec r+\vec r_0)=\vec{A}(\vec r)
+\frac{1}{e^*}\vec\nabla\chi$ 
with
$\chi(\vec r)=(e^*B/2)(\vec r_0 \times \vec r)$
for arbitrary $\vec r_0$,
and the wavefunction transformation 
\beq
\Psi(\vec r)\to\Psi(\vec r -\vec r_0)\,\,
e^{i \frac{e^*B}{2}( \vec r_0 \times \vec r)}\,.
\label{transf}
\eeq
In terms of the holomorphic coordinates,
the  transformation reads for arbitrary $\zeta_0$
\bea \Psi (\zeta,\bar\zeta)
\to
\Psi(\zeta-\zeta_0,\bar\zeta-\bar\zeta_0)
\exp[ \frac{1}{2}(\zeta\bar\zeta_0 - \bar\zeta\zeta_0)]
\,,
\eea
or in the Fourier transformed space 
\bea \tilde\Psi (k,\bar k)
\to
\tilde\Psi(k-k_0,\bar k-\bar k_0)
\exp[ \frac{1}{2}(k\bar k_0 - \bar k k_0)]
\eea
for arbitrary complex momentum $k_0$.
The newly introduced phase factor $v(\{ p_i ,\bar p_i\}) $
in NC quartic term (\ref{k4}) is invariant 
under the gauge transformation $p\to p+p_0$, 
\bea
&&(p_1 \bar p_3 - \bar p_1 p_3)
+ (p_2 \bar p_4 - \bar p_2 p_4) \\
&&\to  (p_1 \bar p_3 - \bar p_1 p_3)
+ (p_2 \bar p_4 - \bar p_2 p_4)
+  p_0 (\bar p_1+\bar p_2-\bar p_3-\bar p_4)
-\bar p_0 (p_1+p_2-p_3-p_4)\,,
\eea
since the  extra contribution proportional to $p_0$ 
( and to $\bar p_0$) vanishes 
due to the momentum conservation. 

There is an other possibility of 
the quartic term, $|\Psi\star\Psi|^2$.
Indeed, the renormalizability of the theory with two
terms are investigated in \cite{c-scalar}.
However, we ruled out the term $|\Psi\star\Psi|^2$
because this term is not invariant under 
the wavefunction transformation (\ref{transf}).
Then, there arises a question: 
how much will the gauge invariant phase factor
$v(\{p_i,\bar p_i\}) $ affect the correlation.  
This issue will be answered in the next section.

\vskip 0.5cm
{\it 3. Equivalence relation:}
The effect of thermal fluctuations in the vortex liquid system
is studied by the partition function
\beq
Z = \int{\cal D}\phi{\cal D} \phi^\dagger \,\,
e^{ - H[\phi,\phi^\dagger]/k_{\!B} T } \,.
\label{Z}
\eeq
One can perform the perturbative calculation using the  
bare propagator obtained from the quadratic term of 
the Hamiltonian $H_2 = K_2$
\beq
\langle\phi^\dagger \,(i\bar k_1)\phi
(-ik_2)\rangle_0  = \frac{1}{2\pi\alpha_2}e^{\bar k_1 k_2}\,.
\eeq  
The $2n$-point function 
$G( k_1 \cdots k_n; k_{n+1} \cdots k_{2n}) = 
\langle\phi^\dagger(i\bar k_1) \cdots \phi^\dagger(i\bar k_n)
\phi(-ik_{n+1}) \cdots \phi(-ik_{2n})\rangle $ 
and its higher order corrections  
are evaluated once the lowest order of the 4-point function is known. 
From $K_4$ (\ref{k4}) the four-point function at the tree level 
is given as
\beaq
G_{\rm NC}^0(k_1, k_2; k_3, k_4) 
&\equiv& 
\begin{picture}(100,40)(0,40)
\put(10,70){\line(1,0){80}}
\multiput(50,12)(0,3){20}{\circle*{2}}
\put(10,10){\line(1,0){80}}
\put(25,7.5){\small $>$}
\put(25,67.5){\small$>$}
\put(70,7.5){\small $>$}
\put(70,67.5){\small$>$}
\put(-5,70){$\bar k_1$}
\put(-5,10){$\bar k_2$}
\put(95,70){$k_3$}
\put(95,10){$k_4$}
\end{picture}
\nonumber\\
\nonumber\\
\nonumber\\
\nonumber\\
&=&-\frac{\alpha_4}{\alpha^4_2}(\frac{1}{8\pi^2})
\exp\big[\bar k_1 k_3 +\bar k_2 k_4\big]
\int\frac{d^2 p}
{(2\pi)^2}\; \tilde g_0 ^{\rm NC}(p,\bar p) \nonumber\\
&& \qquad \times 
\exp\big[-\frac{17}{32} p\bar p
+\frac{5}{8}\bar p (k_3 -k_4) 
- \frac{5}{8} p (\bar k_1 - \bar k_2)\big] .
\label{g4nc}
\eeaq
Rewriting this using 
$\tilde f^{\rm NC}(\vec k)\equiv  \exp\big[-\frac 9{64}k\bar k\Big] \,
\tilde g^{\rm NC}(\vec k)$ we have
\beaq
G^0_{\rm NC}(k_1, k_2; k_3, k_4) 
&=&-\frac{\alpha_4}{\alpha^4_2} \Big(\frac{1}{8\pi^2}\Big)
\exp\big[\bar k_1 k_3 +\bar k_2 k_4\big]
\int\frac{d^2 p}
{(2\pi)^2}\; \tilde f_0 ^{\rm NC}(p,\bar p) \nonumber\\
&& \qquad \times 
\exp\big[-\frac{25}{64} p\bar p
+\frac{5}{8}\bar p (k_3 -k_4) 
- \frac{5}{8} p (\bar k_1 - \bar k_2)\big] .
\eeaq

This is compared with the BNT  case;
\beaq 
G^0_{\rm BNT}(k_1, k_2; k_3, k_4)
&=&-\frac{\alpha_4} {\alpha^4_2}
\Big(\frac{1}{8\pi^2} \Big) 
\exp \Big[\, \bar k_1 k_3 + \bar k_2 k_4 \,\Big] 
\int\frac{d^2 p}{(2\pi)^2}\; 
\tilde g_0^{\rm BNT}(p,\bar p) 
\nonumber\\ 
&&\qquad \times  \exp\big[-\frac {p\bar p}{2}
+\frac{1}{2}\bar p (k_3 -k_4)  
- \frac{1}{2} p (\bar k_1 - \bar k_2)\big]
\nonumber\\ 
&=&-\frac{\alpha_4}{\alpha^4_2} 
\Big(\frac{1}{8\pi^2} \Big) 
\exp\Big[\,\bar k_1 k_3 +\bar k_2 k_4\,\Big] 
\int\frac{d^2 p}{(2\pi)^2}\; 
\tilde f_0^{\rm BNT}(p,\bar p) 
\nonumber\\ 
&&\qquad \times  \exp\Big[-\frac {p\bar p}{4}
+\frac{1}{2}\bar p (k_3 -k_4)  
- \frac{1}{2} p (\bar k_1 - \bar k_2)\Big] . 
\label{g4bnt} 
\eeaq 
where  $\tilde f^{\rm BNT}
\equiv\exp\Big[-\frac 1 4 k\bar k\Big] 
\tilde g^{\rm BNT} (\vec k)$.
Comparing  (\ref{g4nc}) with (\ref{g4bnt})  
we have an equivalence relation if we put 
\beq
\widetilde f^{\rm BNT} (k,\bar k) 
=\Big(\frac{16}{25} \Big)\, \widetilde f^{\rm NC} 
\Big(\frac{4}{5}k,\frac{4}{5}\bar k \Big)\,. 
\label{equiv}
\eeq 
The same relation holds for all order of the perturbation.

\vskip 0.5cm
{\it 4. Remarks and conclusion:}  
We reformulated the lowest Landau level   
effective Ginzburg-Landau theory from the 
non-commutative field theory point of view.  
This NC theory naturally incorporates 
the non-commuting nature of  
coordinates. 
As the consequence of 
the non-local behavior of  
the system due to the phase space reduction,
gauge invariant factor 
$v(\{ p_i ,\bar p_i\}) $ (\ref{v})
appears in the quartic interaction.

The appearance of the new gauge invariant factor 
seems not introduce any new physics 
since there exists the non-trivial equivalence relation (\ref{equiv}) 
between the coupling of NC and that of BNT. 
Does this equivalence relation demonstrate the 
nonrelevance of NC field approach to LLL vortex system? 
The answer is no. 
This is the disguise of the benefit of 
the momentum space description.

Note that NC theory has the bare  function  
$\tilde g^{\rm NC}_0 (k,\bar k)=1$ or  
$ \tilde f^{\rm NC}_0 (k,\bar k)
= e^{-\frac{9}{64}k\bar k}$  and the corresponding bare function for BNT theory 
is given as 
\beq \tilde f^{\rm BNT}_0 (k,\bar k)  
=(\frac{16}{25})\, \widetilde f^{\rm NC}_0 
(\frac{4}{5}k,\frac{4}{5}\bar k) 
= \frac{16}{25} \,  
\exp(-\frac{9}{100}k\bar k), 
\eeq 
or in terms of the $g$ function,  
\beq 
\tilde 
g^{\rm BNT}_0 (k,\bar k) 
=\frac{16}{25}\exp(\frac{4}{25} k\bar k). 
\eeq 
This effective BNT bare coupling cannot be 
Fourier transformed to  the real space  
though it can be formally put into a non-local form in the real space. 
In this sense, the NC theory covers the larger domain of 
coupling constants in coordinate representation
where BNT theory becomes unphysical.  
The gauge invariant phase factor  $v(\{p_i,\bar p_i\})$ 
coming from the NC  field theoretical consideration  
has a very unexpected role  from BNT point of view.   
We note by passing that the specific scaling factor $5/4$ comes 
from the special form of $v(\{p_i,\bar p_i\})$  due to the Moyal product. 
In general one may introduce the arbitrary power of $v$ 
without destroying the gauge invariance and change the scaling factor 
by the same power. 

On the other hand, as far as the RG flow concerns, 
the runaway picture of  $g(\zeta, \bar \zeta)$ does not 
change and hence, signals the first order phase transition.  
This can be seen as follows. 
BNT theory has the bare  function  
$\tilde g^{\rm BNT}_0 (k,\bar k)=1$ 
or  $ \tilde f^{\rm BNT}_0 (k,\bar k)= e^{-\frac{1}{4}k\bar k}$.
The corresponding bare function  of NC theory is given as 
\beq
 \tilde f^{\rm NC}_0 (k,\bar k)  
=(\frac{25}{16})\, \widetilde f^{\rm BNT}_0 
(\frac{5}{4}k,\frac{5}{4}\bar k)
 = \frac{25}{16} \, \exp(-\frac{25}{64}k\bar k), 
\eeq 
or in terms of the $g$ function,  
\beq 
\tilde g^{\rm NC}_0 (k,\bar k)
=\frac{25}{16}\exp(-\frac14 k\bar k),\quad\quad 
g^{\rm NC}_0(\zeta,\bar\zeta)
=\frac{25}{16\pi} \exp(-\zeta\bar\zeta)\,. 
\eeq  
One can follow the one-loop RG analysis in \cite{bnt,mn}
using the equivalent form (\ref{equiv})
and arrive at the same conclusion
since the RG flow shares the same structure 
but with a different initial condition.

Finally, one may study the vortex lattice formation
using the NC formalism.
Minimizing the free energy is equivalent to minimizing the Abrikosov ratio 
\cite{abrikosov},
which, in the NC theory, is given by
\beaq
\beta^{NC}_A= \frac1A \int d^2r (\Psi \star \Psi^\dagger)^2
\Big/
\Big\{ \frac1A \int d^2r |\Psi|^2\Big\}^2 \,
\eeaq
where $A$ is the area of the two-dimensional space. The vortex lattice 
solution satisfies
the periodicity condition,
\beaq
|\Psi(\vec r + \vec r_I )| = |\Psi(\vec r )|\,,
\quad
|\Psi(\vec r + \vec r_{II} )| = |\Psi(\vec r )|
\eeaq
with arbitrary periodicity vectors parametrized by $\vec r_I =\ell^\prime 
(1,0)$
and $\vec r_{II} =\ell^\prime (\nu, \eta)$. The flux quantization condition
gives the area of the unit cell
$\eta \ell^{\prime 2} = 2\pi \ell^2$ with the magnetic length
$\ell$ set to unity in this paper.
In terms of the reciprocal lattice vector $\vec G_{mn}$,
which satisfies
$\vec G_{mn} \cdot \vec r_I = 2\pi m$ and
$\vec G_{mn} \cdot \vec r_{II} = 2\pi n$,
one may put
\beq
\beta_A^{NC} =\frac {16}{25}
\sum_{\vec G_{mn}} e^{- \frac9{100}|\vec G|^2}\,.
\eeq
The minimum value is achieved for a triangular lattice with 
$\beta^{NC}_A=1.7782$.
This is, however, only slightly lower than that for
the square lattice, $\beta^{NC}_A=1.7789$.
In the conventional GL theory, the Abrikosov ratio can be written as
\beq
\beta_A^{BNT} =
\sum_{\vec G_{mn}} e^{- \frac14 |\vec G|^2}\,
\eeq
which gives the well known results, $\beta^{BNT}_A=1.1596$ for a traingular 
lattice
and $\beta^{BNT}_A=1.1803$ for a square lattice.
We note that this expression
can also be obtained if one rescales the NC result by $G \to \frac 54 G$.
From this, one may conclude that a triangular vortex lattice will also be 
formed
in the NC theory, although the difference in free energy between various 
vortex lattice
structures is small compared to that in the conventional GL theory.

In summary, we have studied the vortex system restricted in the LLL using the new
Ginzburg-Landau model inspired by the NC field theory. The quartic term in the new GL model
differs from the conventional one by the gauge invariant phase factor. We have shown that
the effect of this phase factor is the nontrivial rescaling in the correlation functions for the vortex
liquids as well as in the mean field quantities describing the vortex lattice.
This is, at first sight, quite puzzling,
since one cannot rewrite the quartic term (\ref{k4}) in any simpler way by rescaling the momenta
into the form without the phase factor $v$. However, once this phase factor is put into
Gaussian integrations to calculate the correlation functions, 
they produce the nontrivial rescaling as we have found
in this article. It will be interesting to study the possibility of the physical quantity describing
the vortex liquids, for which the phase factor
produces other effects than the rescaling. This is left to future work.

\vskip .5cm
{\it Acknowledgements:}
This work was supported by grant No.\ R01-1999-00018 from the
interdisciplinary research program of the KOSEF.
JY and CR  would like to acknowledge the hospitality
from the YVRC, CR from KIAS and SPhT, and VP 
from KIAS.

\end{document}